\documentclass[12pt]{article}

\def\thefootnote{\fnsymbol{footnote}}
\def\11{\mbox{$1$}}

\renewcommand{\thefootnote}{\alph{footnote}}

\newcommand{\rref}[1]{(\ref{#1})}
\newcommand{\beqn}{\begin{equation}}
\newcommand{\eeqn}{\end{equation}}
\newcommand{\beqarr}{\begin{eqnarray}}
\newcommand{\eeqarr}{\end{eqnarray}}

\newcommand{\matc}{\begin{array}{c}}
\newcommand{\matcc}{\begin{array}{cc}}
\newcommand{\matccc}{\begin{array}{ccc}}
\newcommand{\matcccc}{\begin{array}{cccc}}
\newcommand{\emat}{\end{array}}

\newcommand{\IH}{\relax{\rm I\kern-.18em H}}
\newcommand{\IR}{\relax{\rm I\kern-.18em R}}
\newcommand{\IK}{\relax{\rm I\kern-.18em K}}
\newcommand{\II}{\hbox{\rm 1\kern-.28em I}}
\newcommand{\Is}{\relax{\rm 1\kern-.35em 1}}

\begin{document}

\begin{titlepage}

February 2001         \hfill
\vskip -0.55cm
\hfill  RU-01-2-B
\\
\vskip -0.5cm
\hfill  GSUC-CUNY-PHYS-01-3
\\
\vskip -0.5cm
\hfill  CU-TP-1007
\begin{center}
\vskip .15in
\renewcommand{\thefootnote}{\fnsymbol{footnote}}
{\large \bf Quantum Mechanics on the Noncommutative Torus}
\vskip .25in
Bogdan Morariu$^{a}$ and
Alexios P. Polychronakos$^{b,}$\footnote{On leave from
Theoretical Physics Dept., Uppsala University,
  Sweden and Physics Dept., University of Ioannina, Greece}
\vskip .25in

{\em    $^{a,\,b}$Department of Physics,        Rockefeller University   \\
        New York, NY 10021      \\
\vskip .25in
        $^{b}$Department of Physics, City College of the CUNY  \\
        New York, NY 10031    \\
\vskip .25in
        $^{b}$Department of Physics, Columbia University   \\
        New York, NY 10027}
\end{center}
\vskip .25in
\vskip .05in {\rm E-mail: morariu@summit.rockefeller.edu, poly@teorfys.uu.se}
\begin{abstract}
We analyze the algebra of observables of a charged
particle on a noncommutative torus in a constant magnetic field.
We present a set of generators of this
algebra which coincide with the generators for a commutative torus but
at a different value of the magnetic field, and demonstrate the existence of a
critical value of the magnetic field for which the algebra reduces. We then
obtain the irreducible representations of the algebra and relate them to
noncommutative bundles. Finally we comment on Landau levels,
density of states and the critical case.
\end{abstract}

\end{titlepage}

\newpage
\renewcommand{\thepage}{\arabic{page}}

\setcounter{page}{1}
\setcounter{footnote}{0}

\section{Introduction}
\label{Intro}

Noncommutative spaces arise as solutions of matrix models 
and in the effective description of branes in string theory 
\cite{deWit:1988ig}-\cite{Seiberg:1999vs}. The fluctuations of
these brane or the matrix model solutions are described by
noncommutative field theories. It is expected that quanta of these
field theories will represent particles moving on the underlying
noncommutative spaces. It is therefore of interest to examine the
dynamics of these quantum mechanical particles.

Although noncommutative field theories have been extensively studied,
the corresponding quantum mechanical problem has received relatively
little attention until recently
\cite{Dunne:1990hv}-\cite{Nair:2000ii}. Some related studies
of finite quantum mechanics and its relation to the noncommutative 
torus are~\cite{Athanasiu:1996ni}-\cite{Floratos:2000nr}, and for a
discussion of the modular invariace 
see~\cite{Aldaya:1996zd}\cite{Guerrero:1997fe}.
In \cite{Nair:2000ii}, 
in particular, the problem of a quantum particle moving
on a noncommutative plane and sphere was examined and solved. The
results revealed a Landau level picture analogous to the commutative
case. An important qualitative difference, however, was a modified
density of states and the existence of a critical value of the magnetic
field at which this density diverges. This may have a relevance to
the recently proposed analogy between noncommutative field theory
and the quantum Hall effect \cite{Susskind:2001fb}.  

The purpose of this paper is to analyze the Landau problem for the
case of a flat periodic space, that is, a two-torus. This situation is
interesting even in the commutative case, being closely related to the
Hofstadter problem. As we will show, a mapping can be established between
the commutative and noncommutative cases, revealing the features of the
model and the emergence of the critical magnetic field.

We will follow the approach of identifying the algebra of physical 
observables for the model and finding its irreducible representations.
This is conceptually more fundamental than corresponding treatments 
based on explicit wave equations and bypasses the questions of extended
gauge invariance, single-valuedness of the wavefunction etc. It is similar
in spirit to the approach of isolating and examining only the gauge 
invariant observables and their algebra, rather than working with 
a gauge theory and imposing gauge invariance on the states. A connection
with more traditional approaches will also be given.

The structure of the paper is as follows. In Section~\ref{Algebra} we write
the algebra of observables on the noncommutative torus and establish its
equivalence with a modified commutative torus. In Section~\ref{Commutative}
we use this equivalence and obtain irreducible representations of
the algebra. In Section~\ref{NC}
we review the algebra of observables and its representations for the
noncommutative plane. In Section~\ref{NCB} we recover the torus
representations through a reduction of the planar representation and show
that they correspond to quantum bundles over the noncommutative torus.
In the last section we discuss the energy eigenstates (Landau levels)
and the density of states and comment on the critical case.

\section{Algebra of observables}
\label{Algebra}

A noncommutative plane is defined in terms of two flat noncommutative
coordinates $X_1 , X_2$ obeying the commutation relation
\beqn
[ X_1 , X_2 ] = i \Theta~,
\label{ncx}
\eeqn
with $\Theta$ a c-number parameter, equal to the noncommutative
length scale squared.
The motion of a particle on such a plane will be described by the
above noncommutative coordinates and two corresponding momenta $P_i$,
defined as shift operators on the $X_i$. In the presence of a constant
magnetic field $\cal B$ the commutator of the $P_i$ does not vanish and
becomes proportional to $\cal B$. ($P_i$ are the `gauge invariant' or
`kinematical' momenta, rather than the `canonical' momenta.) The complete
algebra of observables is, then
\beqn
\left\{~
\begin{array}{ccc}
\left[\,X_1  , X_2 \,\right]&=& i\Theta~, \\
\left[\,X_i , P_j \,\right] &=& i \delta_{ij}~,\\
\left[\,P_1,P_2\,\right]   &=& i\,{\cal B}~.
\end{array}\right.
\label{Plane}
\eeqn
The Hamiltonian is the free one
\beqn
H=\frac{1}{2}\,(P_1^2+P_2^2)~.
\label{Ham}
\eeqn

To describe a particle on a noncommutative torus, we further impose
the periodicity condition that
\beqn
{\vec X} \sim {\vec X} + {\vec a}_i ~,~~~i=1,2~,
\label{period}
\eeqn
where the components of $ {\vec a}_i$ are $c$-numbers.
This represents an oblique torus with period vectors ${\vec a}_1 ,
{\vec a}_2$ and area $A = {\vec a}_1 \times {\vec a}_2 = a_{11} a_{22}
- a_{12} a_{21}$.

Clearly the
$X_i$ are not physical operators, since they are not uniquely fixed by
the position of the particle on the torus. As physical operators we
take the exponentials
\[
U_i = e^{i {\vec b}_i \cdot {\vec X} }~,
\]
where ${\vec b}_i$ are the dual torus vectors satisfying
\[
{\vec a}_i \cdot {\vec b}_j = 2\pi \delta_{ij}~.
\]
The above unitary $U_i$ are invariant under the shifts~\rref{period}
and are, therefore, physical operators. The complete
set of physical observables for this particle are the two unitary
`position' operators $U_i$ and the two hermitian momenta $P_i$
satisfying the commutation relations
\beqn
\left\{~
\begin{array}{ccc}
U_1 U_2~~ &=& U_2 U_1~ e^{-i\theta}~,\nonumber  \\
P_i \, U_j ~~&=& U_j \, (P_i+a_{ji})~, \\
\left[\,P_1,P_2\,\right]   &=& i\,{\cal B}~,\nonumber
\end{array}\right.
\label{alga}
\eeqn
with $\theta = (2\pi)^2 \Theta / A$ a dimensionless parameter
($1/\theta$ effectively counts the noncommutative area `quanta' contained
in the torus). Finally, we can cast (\ref{alga}) in a form not explicitly
involving the period vectors by defining new momenta
\[
p_i = \frac{1}{2\pi} {\vec a}_i \cdot {\vec P}~,
\]
in terms of which we obtain
\beqn
\left\{~
\begin{array}{ccc}
U_1 U_2~~ &=& U_2 U_1~ e^{-i\theta}~, \\
p_i \, U_j ~~&=& U_j \, (p_i+\delta_{ij})~, \\
\left[ \, p_1, p_2 \,\right]   &=& i\,{B}~,
\end{array}\right.
\label{alg}
\eeqn
where
\[
B = \frac{ A {\cal B}}{(2\pi)^2}~.
\]
The associative algebra generated by $U_i$ and $p_i$ and satisfying
the relations~\rref{alg} will be denoted by ${\cal T}_{\theta,B}$\,.
We  can also introduce $x_i = {\vec b}_i \cdot {\vec X}$
and rewrite~\rref{Plane} as
\beqn
\left\{~
\begin{array}{ccc}
\left[\,x_1  , x_2 \,\right]&=& i\theta~, \\
\left[\,x_i , p_j \,\right] &=& i \delta_{ij}~,\\
\left[\,p_1,p_2\,\right]   &=& i\,{B}~.
\end{array}\right.
\label{plane}
\eeqn
The associative algebra generated by $x_i$ and $p_i$ and satisfying
the relations~\rref{plane} will be denoted by ${\cal P}_{\theta,B}$\,.
Note that the torus algebra  ${\cal T}_{\theta,B}$\, is a subalgebra of
the plane algebra ${\cal P}_{\theta,B}$\,.
The Hamiltonian is given by
\[
H=\frac{1}{2}\,(P_1^2+P_2^2) = \frac{b_1^2}{2} p_1^2 +
\frac{b_2^2}{2} p_2^2 + {\vec b}_1 \cdot {\vec b}_2
( p_1 p_2 + p_2 p_1 )~.
\]

We will now show that ${\cal T}_{\theta,B}$ is isomorphic to
${\cal T}_{0,\tilde{B}}$ where $\tilde{B}=B/(1-B\theta)$\,. First let
\[
\tilde{U}_{1}\equiv U_1 e^{i\alpha  p_2}~,~
\tilde{U}_{2}\equiv U_2 e^{i\beta  p_1} ~,
\]
where $\alpha$ and $\beta$ are two real $c$\,-numbers. Then the
$\tilde{U}_i$'s commute if
\beqn
\alpha \beta {B} +\alpha -\beta - \theta=0~.
\label{rel}
\eeqn
One can also show that
\[
\left\{~
\begin{array}{ccc}
p_1 \tilde{U}_1 &=& \tilde{U}_1(p_1 +1-\alpha {B})~,
\nonumber \\
p_2 \tilde{U}_2 &=& \tilde{U}_2(p_2 +1+\beta {B})~.
\nonumber
\end{array}\right.
\]
If we set $1-\alpha {B}=\pm(1+\beta {B})$ and
define $\tilde{p}_1 =p_1/(1-\alpha {B})$ and $\tilde{p}_2 =
\pm p_2/(1-\alpha {B})$ we have
\[
\left\{~
\begin{array}{ccc}
\left[ \tilde{U}_1,\tilde{U}_2 \right] &=&0~,\nonumber\\
\tilde{p}_i \tilde{U}_j~~ &=& \tilde{U}_j(\tilde{p}_i+\delta_{ij})~,\nonumber\\
\left[ \tilde{p}_1,\tilde{p}_2 \right] ~&=&i{\tilde B}~,\nonumber
\end{array}\right.
\]
where
\[
{\tilde B}= \pm \frac{{B}}{(1-\alpha
  {B})^2}~.
\]
Using~\rref{rel} one can show that $(1-\alpha {B})^2= \pm
(1-\theta B)$,
thus the choice of sign is dictated by the sign of $1-\theta B
= 1 - \Theta{\cal B} $\,. Then
$\tilde{U}_i$ and $\tilde{p}_i$ generate
the algebra of observable  ${\cal T}_{0,\tilde{B}}$\,
of a charged particle in a magnetic field $\tilde{B}$
on a commutative torus.
In terms of $\tilde{p}_i$\, the Hamiltonian takes the form
\[
H=\frac{|1-\Theta{\cal B}|}{2}\left\{ \frac{b_1^2}{2} \tilde{p}_1^2 +
\frac{b_2^2}{2} \tilde{p}_2^2 + {\vec b}_1 \cdot {\vec b}_2
( \tilde{p}_1 \tilde{p}_2 + \tilde{p}_2 \tilde{p}_1 ) \right\}~,
\]
thus it only differs from the standard Hamiltonian on the commutative
torus by an overall normalization.

When $1-\Theta {\cal B}$\, vanishes we cannot
define $\tilde{p}_i$ as above. For the choice $\alpha = -\beta
=\theta$, however,
we note that each $p_i$ commutes with
the $\tilde{U}_i$'s which mutually commute (this was not possible before).
Thus for the critical value ${\cal B}=\Theta^{-1}$ the algebra
reduces into the direct
product of a Heisenberg algebra and two commuting U(1) operators
and becomes
\beqn
\left\{~
\begin{array}{ccc}
\tilde{U}_1 \tilde{U}_2~~ &=& \tilde{U}_2 \tilde{U}_1~,\nonumber \\
 \left[\,p_i ,\tilde{U}_j \,\right]&=& 0~, \label{Crit}\\
  \left[\,p_1,p_2\,\right]   &=& i\,{B}~.\nonumber
\end{array}\right.
\eeqn
This signals the reduction of the Hilbert space at criticality.

In the next sections we use two different methods to study the
representations of the algebra ${\cal T}_{\theta,B}$, first using the
equivalence to the commutative torus, and then obtaining the quantum bundles
over the noncommutative torus.

\section{Irreducible representations}
\label{Commutative}

Since the noncommutative algebra (away from criticality) is equivalent to the
commutative one with a new value of the magnetic field $\tilde B$, it suffices
to study the irreducible representations of the algebra~\rref{alg}
at $\theta=0$. We need the Casimirs of the algebra. First note that
the operators
\[
\left\{~
\begin{array}{ccc}
W_1&=& \exp (-i \frac{m_1 }B \tilde{p}_1 ) \tilde{U}_2^{~m_1}~, \\
W_2&=& \exp (-i \frac{m_2 }B \tilde{p}_2 ) \tilde{U}_1^{-m_2}~,
\end{array}\right.
\]
where $m_1$ and $m_2$ are arbitrary integers, commute with $\tilde{p}_i$.
They will also commute with $\tilde{U}_i$ if
\beqn
2\pi \tilde{B}= m/n \label{Bquant}
\eeqn
for some integers $m$ and $n>0$, and we take both $m_1$ and $m_2$ to be
multiples of $m$. If we take $(m,n)$ to be relatively prime integers, we obtain
the minimal operators $W_i$, forming a complete set of generators of
the Casimirs of the algebra, by choosing $m_1 = m_2 = m$. So we have
\[
\left\{~
\begin{array}{ccc}
W_1&=& e^{-2\pi i n \tilde{p}_1 } \tilde{U}_2^{~m}~, \\
W_2&=& e^{-2\pi i n \tilde{p}_2 } \tilde{U}_1^{-m}~.
\end{array}\right.
\]
For $n=1$, \rref{Bquant} is the familiar condition for an integer
number of magnetic flux
quanta through the surface of the torus, but with the modified
magnetic field $\tilde B$ now entering the quantization condition. The
representation
of the algebra of observables in that case is rather straightforward.
As we will see, for $n > 1$ one can
obtain a representation by either considering an enlarged torus of area
$n A$ or equivalently by introducing an internal quantum number
corresponding to an $n$-fold wave function.

Next we find a complete set of commuting operators for the algebra containing
${\tilde U}_i$\,. The most general choice (up to some ${\tilde U}_i$
factors) is to add the following two operators
\[
Z_i = e^{-2\pi i \vec{N}_{i} \cdot \vec{\tilde{p}}} ~,~i=1,2
\]
defined in terms of two arbitrary integral vectors $\vec{N}_1$ and $\vec{N}_2$
satisfying $n=\vec{N}_1\times \vec{N}_1 = N_{11} N_{22} -N_{12}
N_{21}$\,. (Note that  $\vec{N}_1 , \vec{N}_2$ define an $n$-fold
enlarged torus.)
The set $\{\tilde{U}_1, \tilde{U}_2, Z_1, Z_2  \}$ is complete and in
particular the Casimir operators $W_i$ can be written in terms of the
elements of this set
\beqn
\left\{~
\begin{array}{ccc}
W_1&=& e^{\pi i m N_{22} N_{12}} Z_1^{N_{22}} Z_2^{-N_{12}} \tilde{U}_2^{~m}~,
\label{Cas}\\
W_2&=&
e^{\pi i m N_{11} N_{21}} Z_1^{-N_{21}} Z_2^{N_{11}}\tilde{U}_1^{-m}~.
\nonumber
\end{array}\right.
\eeqn

Then we can find a state denoted $|\vec{0}\rangle$ which
satisfies\footnote{We could start  with an arbitrary $U_i$ eigenstate
and use exponentials of $p_i$ to get the state  $|\vec{0}\rangle$ which
has  $\tilde{U}_i$ eigenvalues $1$.}
\beqn
\left\{~
\begin{array}{ccc}
\tilde{U}_i |\vec{0}\rangle&=&|\vec{0}\rangle~, \label{U0}\\
Z_i |\vec{0}\rangle&=&e^{i\zeta_i}|\vec{0}\rangle ~.\nonumber
\end{array}\right.
\eeqn
An irreducible representation is obtained by acting on this state
with operators which do not commute with $\tilde{U}_i$ or $Z_i$\, and modding
out by zero norm states.  The following states
\beqn
|\vec{\phi}\rangle = e^{-i\vec{\phi}\cdot\vec{\tilde{p}}} |\vec{0}\rangle~
\label{def}
\eeqn
are obtained in this way and satisfy
\[
\left\{~
\begin{array}{ccc}
\tilde{U}_i |\vec{\phi}\rangle&=&e^{i\phi_i}|\vec{\phi}\rangle~, \\
Z_i
|\vec{\phi}\rangle&=&e^{i(\zeta_i-\frac{m}{n}
\vec{N}_{i}\times \vec{\phi})}
|\vec{\phi}\rangle~.
\end{array}\right.
\]
Note that the states $|\vec{\phi}\rangle$ and $|\vec{\phi}\,'\rangle$
have the same
$U_i$ and $Z_i$ eigenvalues if
\[
\vec{\phi}\,' =\vec{\phi}+2\pi \vec{N}_j~,
\]
and need not be linearly independent. In fact, it is consistent to
identify them, as
\beqn
|\vec{\phi}+2\pi \vec{N}_i\rangle
=
e^{i(\zeta_i-\frac{m}{2n}\vec{N}_{i}\times \vec{\phi})}|\vec{\phi}\rangle~.
\label{PERIOD}
\eeqn
In other words, the difference of the above two states is a null state and can
be consistently set to zero.
The independent states are thus labeled by the vector $\vec{\phi}$ living in
a fundamental cell of the enlarged lattice generated by $\vec{N}_i$
and they form
an irreducible representation. Using~\rref{Cas} we obtain the following
expressions for the Casimirs $W_i = e^{i\omega_i}$ in this representation:
\[
\omega_i =  M_{ij} \,
\zeta_j +  \pi l_j ~,
\]
where $l_i= m M_{i1} M_{i2} {\rm mod}(2)$\, (no $i$\, summation) and $M$ 
is the integral matrix satisfying $MN=n \II$\,. We note that the Casimirs
of the representation are related to the Wilson lines around the periods
of the torus.

We have shown that for every choice of an integral lattice of area $n$
and of phases $\zeta_i$, we can construct in a standard way an
irreducible representation of the algebra ${\cal T}_{0,\tilde B}$
starting from an anchor state satisfying~\rref{U0}.
We will now show that any two such representations (with
different $\vec{N}_{i}$
and $\zeta_i$) having the same Casimirs are equivalent.

First we give an abstract proof.
Consider two representations: the first obtained using $\vec{N}_i$ and
$\zeta_i$ and the second  obtained using $\vec{N}'_i$ and
$\zeta'_i$\,. Since the operators $Z_i$ also act in the second
representation and they commute with $\tilde{U}_i$ they can be simultaneously
diagonalized. Taking an arbitrary eigenvector and acting on it with
translation operators we can obtain an eigenvector with
unit $\tilde{U}_i$ eigenvalues. The result is also an eigenvector
of $Z_i$\, with some
eigenvalues $e^{i\zeta_i}$. Thus we have found an anchor state
of the first representation as a state in the second representation
and thus the two representations are equivalent.

We can also work out the explicit map between states in two
representations. First we find an anchor state of the first
representation (denoted with unprimed states) as a state in the second
representation (denoted with primed states).
Since $Z_i$ are lattice translation operators we make the
following ansatz:
\beqn
 |\vec{0}\rangle =\sum_{k_1,k_2=0}^{n-1}
C_{k_1 ,k_2} | 2\pi k_i \vec{N}_{i}\rangle'~.
\label{anchor}
\eeqn
On the right hand side the states are in the primed representation but
the sum is over lattice points of the unprimed representation.
Assuming that we can shift the summation index the eigenvalue equations
for $Z_i$ give the following equations for the coefficients
\beqn
\left\{~
\begin{array}{ccc}
C_{k_1+1 , k_2} &=& e^{-i\zeta_1}\, e^{-\pi i m k_2} C_{k_1, k_2}~,
\label{Ceq}\\
C_{k_1 , k_2+1} &=& e^{-i\zeta_2}\, e^{+\pi i m k_1}  C_{k_1, k_2}~.\nonumber
\end{array}\right.
\eeqn
These are solved by
\[
C_{k_1 k_2} =
e^{- i \vec{k} \cdot \vec{\zeta}}\,
e^{\pi i m k_1 k_2}~,
\]
for which one can check that we have a periodic summand
in~\rref{anchor} if the phases $\zeta_i$ and $\zeta'_i$ and the two
lattices are such that they give the same Casimirs~\rref{Cas}.
By acting with $e^{- i\vec{\phi}\cdot\vec{p}}$ on the anchor state we
obtain
\[
|\vec{\phi}\rangle =
\sum_{k_1,k_2=0}^{n-1}
e^{- i \vec{k} \cdot \vec{\zeta}}\,
e^{\pi i m k_1 k_2}
e^{i\frac{m}{2n}  k_i \vec{N}_{i}\times\vec{\phi}}
| \vec{\phi}+2\pi k_i \vec{N}_{i}\rangle'~.
\]
which establishes the complete explicit mapping between the two
representations.

The above concludes the derivation of the irreducible Hilbert space for the
particle on the torus. An important point is that the position
operators ${\tilde U}_i$ do not suffice to fully characterize the states;
the additional
operators $Z_i$ are also required. (They are only absent in the case of integer
quantization of $\tilde B$, that is $n=1$, in which case the $Z_i$ can
be expressed
entirely as functions of the ${\tilde U}_i$ and the Casimirs $W_i$.)
This means that
the set of ${\tilde U}_i$ eigenstates alone is not complete and a
`torus wavefunction' description of the states in terms of (quasi-)
periodic functions on the torus is
inadequate. This is common to both the commutative and noncommutative case.
To fully specify the state an additional set of discrete degrees of freedom are
needed. Indeed, looking at~\rref{PERIOD} we see that $\vec \phi$, which labels
independent states, takes values on an $n$-fold enlarged torus. Each
point on the
fundamental torus has $n$ images on the enlarged torus and we need to know the
value of the wavefunction on each of these images to fully specify the state.
This amounts to promoting the ${\tilde U}_i$-wavefunction into an $n$-component
vector.

To make this more explicit, let us consider the representation defined
in terms of the enlarged lattice ${\vec N}_1 = (n,0)$, ${\vec N}_2 = (0,1)$.
This corresponds to enlarging the fundamental torus $n$-fold in the
$1$-direction. There is no loss of generality since representations defined in
terms of any ${\vec N}_i$ are equivalent. Then define the states
\[
|{\vec \phi};q\rangle\kern-.28em\rangle = e^{-\frac{i}{n}(\zeta_1 -
\frac{m}{2n}
\phi_2 )q} |\phi_1 + 2\pi q , \phi_2 \rangle
~,~~ q=0,\dots n-1~.
\]
For each $\vec \phi$ the above states form an $n$-vector with components
labeled by $q$. By virtue of~\rref{PERIOD} this vector is quasiperiodic on
the fundamental torus for the variable $\vec \phi$, namely
\[
\left\{~
\begin{array}{ccc}
|\phi_1 + 2\pi, \phi_2 ; q\rangle\kern-.28em\rangle  &=&e^{\frac{i}{n}(\zeta_1
- \frac{m}{2n} \phi_2 )} |\phi_1, \phi_2 ;
q+1\rangle\kern-.28em\rangle~,
\nonumber\\
|\phi_1 , \phi_2 +2\pi ;q\rangle\kern-.28em\rangle &=& e^{i(\zeta_2
+ \frac{m}{2n} (\phi_1 + 2\pi q)} |\phi_1, \phi_2 ; q+1
\rangle\kern-.28em\rangle~.
\end{array}\right.
\]
We see that shifts in $\phi_1$ and $\phi_2$ act as ``shift'' and ``clock''
matrices on the $q$ components (modulo $\vec \phi$-dependent phases).
So, overall, the large gauge transformations associated with shifts in the
fundamental torus for $\vec \phi$ have been promoted to $U(n)$ nonabelian
transformations.

\section{Particle on the Quantum Plane in B-field}
\label{NC}

An alternative way of obtaining irreducible representation of the
algebra of observables~\rref{alg} of the charged particle on the
quantum {\em torus}
is to start with representations of the algebra of
observables~\rref{plane}\,
of the particle on the quantum {\em plane}. These decompose as direct sums of
irreducible representations of the algebra~\rref{alg}.
In this section we present two ways of obtaining representations of the
algebra of observable in the quantum plane case~\rref{plane}\,,
and then show that they are equivalent. In the next section we show
how to select a particular irreducible representation of the
algebra ~\rref{alg}.

The first method is similar to what we did in the previous chapter
for the quantum torus. First we introduce a new set of
generators
\[
\left\{~
\begin{array}{ccc}
{\tilde x}_i &=& x_i -\kappa \varepsilon_{ij} p_j~,\\
{\tilde p}_i &=& \frac{p_i}{1+\kappa B}~,
\end{array}\right.
\]
where $x_i = {\vec b}_i \cdot {\vec X}$ and
\[
\kappa =\frac{(1-\theta B)^{1/2}-1}{B}~.
\]
They satisfy
\beqn
\left\{~
\begin{array}{ccc}
\left[{\tilde x}_1,{\tilde x}_2\right]
 &=&
0  ~, \nonumber\\
\left[{\tilde x}_i,{\tilde p}_j\right]
 &=&
i \delta_{ij}  ~, \label{tilde}\\
\left[{\tilde p}_1,{\tilde p}_2\right]
 &=&
i {\tilde B}  \nonumber~.
\end{array}\right.
\eeqn
The associative algebra generated
by ${\tilde x}_i$ and ${\tilde p}_i$ is identical to the commutative
magnetic algebra
and thus isomorphic to two copies of the Heisenberg algebra\footnote{This can
be shown by making an additional $\tilde x$-dependent linear shift of the
$\tilde{p}_i$\, generators.}. Therefore it has a unique irreducible
representation.  Any state $|f)$ in this
representation can be expanded in ${\tilde x}_i$ eigenstates
\beqn
| f)=\int {d^2 \kern-.18em y} \,\, f(\vec{y}) |\vec{y} )~,
\label{State}
\eeqn
where the eigenstates and their relative phases are chosen to satisfy
\[
\left\{~
\begin{array}{ccc}
{\tilde x}_i |\vec{y}) &=& y_i |\vec{y}) ~,\\
 |\vec{y} )~~&=& e^{-i \vec{y} \cdot {\tilde p}} |\vec{0} )~.
\end{array}\right.
\]
This associates a commutative wavefunction $f(\vec y)$ with each
state. Note that we will use a round bracket for states on the plane and an
angle bracket for states on the torus.

The second method of obtaining the irreducible representation of the
quantum algebra of observables~\rref{plane} is through the $x$-operator
representation of fields on the noncommutative plane. Define the
commuting operators
\beqn
\Delta_i = \frac{p_i + \kappa B\theta^{-1} \varepsilon_{ij} x_j}{1+\kappa B} ~,
\label{Delta}
\eeqn
with $\kappa$ as above. The algebra of observables for the quantum
plane~\rref{plane} is also generated by $x_i$ and $\Delta_i$
which satisfy
\beqn
\left\{~
\begin{array}{ccc}
\left[x_1,x_2\right]
 &=&
i \theta  ~,\nonumber  \\
\left[x_i,\Delta_j\right]
 &=&
i \delta_{ij}  ~,\label{XDelta} \\
\left[\Delta_1,\Delta_2\right]
 &=&
0 ~. \nonumber
\end{array}\right.
\eeqn
One obvious representation of the algebra defined by
relations~\rref{plane} is the algebra
itself with the action given by the left algebra multiplication. The
subalgebra generated by $x_i$ is also a representation. To see this first note
that, since $\Delta_i$ commute, we can define a state $|1)$ satisfying
\beqn
\Delta_i |1) =0~. \label{One}
\eeqn
Then a representation is obtained by acting with arbitrary numbers
of $x_i$ and $p_i$ on $|1)$\,.  We can eliminate the $p_i$\,'s
using the inverse of~\rref{Delta}
\[
p_i = (1+\kappa B) \Delta_i - \kappa B\theta^{-1} \varepsilon_{ij} x_j~,
\]
and after commuting all the $\Delta_i$ to the right
and using~\rref{One} every state can be written as
$\hat{f}|1)$\,,
where $\hat f$ denotes an operator constructed out of $x_i$\,'s.
Thus we can identify this representation
with the associative algebra generated by $x_i$ and it is convenient
to drop the state  $|1)$ and
simply write states as $\hat f$\,. The generators act on such a state as
\[
\left\{~
\begin{array}{ccc}
x_i( \hat f) &=& x_i {\hat f}~, \nonumber \\
\Delta_i ( {\hat f}) &=& [\Delta_i, {\hat f}]
=\theta^{-1} \,\varepsilon_{ij} [x_i, {\hat f}]~,
\end{array}\right.
\]
thus $x_i$  acts by left multiplication and $\Delta_i$ as a
commutator.
In the last line we first used~\rref{One}, then, used the fact that in
commutators we can freely substitute  $\theta^{-1} \,\varepsilon_{ij} x_j$
for $\Delta_i$
to express the result of the action of $\Delta_i$ only in terms of $x_i$\,.

Next we find the explicit state-operator map
relating the two representations. First we
need the operators corresponding to the
eigenstates $|\vec{y})$\,. After
expressing the generators in~\rref{tilde} in terms of the generators
in~\rref{XDelta} we have
\[
\tilde{x_i} (\hat f) = \left( 1+\frac{\kappa}{\theta}\right)
\{ x_i,\hat{f} \}~.
\]
The operator $\hat{f}$ corresponding to the state
$|\vec{0})$ must satisfy
\[
\{ x_i,\hat{f} \} =0~,
\]
thus it must be proportional to the parity operator $P$. A
representation of $P$ is \cite{Gross:2000ph}
\[
P=\frac{1}{(2\pi)^2}\int {d^2 \kern-.18em k} \,\,e^{i\vec{k}\cdot \vec{x}}~.
\]
We can fix the overall normalization by requiring that we have the map
\[
|\vec{0}) \rightarrow P~.
\]
Then we also have
\[
|\vec{y}) \rightarrow e^{-i\vec{y}\cdot \tilde{p}}(P)
=\frac{1}{(2\pi)^2}
\int {d^2 \kern-.18em k} \,\,e^{i\vec{k}\cdot (\vec{x}-\lambda\vec{y})}
~,
\]
where
\[
\lambda=\frac{(1-\theta B)^{1/2}+1}{2(1-\theta B)^{1/2}}=1
+\frac{1}{4}\theta B+\ldots ~.
\]
Finally for an arbitrary state~\rref{State} we have
\beqn
\label{STAR}
| f) \rightarrow {\hat f}=
\frac{1}{(2\pi)^2}
{\int  \kern-.50em\int} {d^2 \kern-.18em y}\,{d^2 \kern-.18em k}
\,\,f(\vec y)
e^{i\vec{k}\cdot (\vec{x}-\lambda\vec{y})} ~.
\eeqn
This is almost the standard map from commutative functions to operators for
the quantum plane except for the factor $\lambda$\,, and it can be
used to define a $*$-product on the space of commutative functions on
the plane.  Note that for
$B=0$\, we have $\lambda=1$ and the map reduces to the standard
case. One can also check that
\[
2\pi \theta {\,\rm Tr}(\hat f)=\int  {d^2 \kern-.18em y} \,\,f(\vec y)~,
\]
thus the inner product of commutative wave functions is mapped into
\[
\left( \hat{f},\hat{g}\right)
=
2\pi \theta {\,\rm Tr}({\hat f}^{\dagger}\hat{g})~.
\]

\section{Quantum Bundles}
\label{NCB}

We now discuss how to formally obtain the representations for the
torus algebra~\rref{alg} by reducing the corresponding representation for the
plane. The magnetic translations
\[
D_i = \frac{1}{1-\theta B} (p_i -B \varepsilon_{ij} x_j )~,~i=1,2~,
\]
commute with $p_i$ and shift the $x_i$ in the standard way. Therefore
the
operators
\[
V_i \equiv e^{2\pi  i D_i}
\]
also commute with $U_i$\,. Thus $V_i$\, generate the commutant
of the torus algebra~\rref{alg} in the planar algebra~\rref{plane}.
The $V_i$, however,
do not mutually commute but rather satisfy the `clock and shift' algebra
\beqn
V_1 V_2 = V_2 V_1\, e^{2\pi i m/n}~.
\label{Valg}
\eeqn
A maximal commutative subalgebra of the algebra generated by $V_i$\,
is generated by the operators
\[
T_i \sim V_1^{N_{i1}} V_2^{N_{i2}}~,
\]
if $\vec{N}_1\times \vec{N}_2=n$\,. We choose the phases such that
\[
T_i \equiv \exp\left(-\frac{2\pi i B}{1-\theta B}\vec{N}_i \times
  \vec{x}\right)
\exp\left(\frac{2\pi i}{1-\theta B}\vec{N}_i\cdot \vec{p}\right)~.
\]
Since $T_i$ commute with the generators of the algebra~\rref{alg} we
can obtain a representation of this algebra by requiring states to
satisfy
\beqn
T_i |f\rangle = e^{-i \zeta_i}|f\rangle~.
\label{TI}
\eeqn
This representation is in fact irreducible. If we suppose that it is
not, there must exist an operator commuting with the generators of the
algebra~\rref{alg} and taking distinct eigenvalues in each irreducible
subrepresentation. But then this operator must commute with $T_i$ and,
since the algebra generated by $T_i$ is maximal,
it must itself be expressed in terms of $T_i$. Thus it must be
proportional to the identity on the whole representation.
Using the constraint~\rref{TI} on a state~\rref{State}
implies the following quasi-periodicity
\beqn
f(\vec{y} + 2\pi \vec{N}_i) = e^{-i(\zeta_i -\frac{m}{2n} \vec{N}_i
  \times \vec{y})}f(\vec{y})~.
\label{c-trans}
\eeqn
Using this quasi-periodicity we can rewrite the integral over the
plane as an integral over a fundamental cell of the lattice generated
by $\vec{N}_i$
\[
|f\rangle =
\int_{\rm cell}
{d^2 \kern-.18em y} \,\, f(\vec{y}) |\vec{y}\rangle~,
\]
where
\[
 |\vec{y}\rangle=
\sum_{k_i} (-1)^{m k_1 k_2}
e^{-i k_i(\zeta_i- \frac{m}{2n} \vec{N}_i\times \vec{x})}
| \vec{y}+2\pi k_i\vec{N}_i) ~.
\]
Then one can check that the states
$|\vec{y}\rangle$ satisfy the
quasi-periodicity~\rref{PERIOD}. Thus we have identified the
representations of Section~\ref{Commutative} embedded in the planar
representation.

Alternatively we can obtain the same representations in the noncommutative
plane operator representation by imposing
\beqn
T_i (\hat f) =e^{-i \zeta_i} (\hat f)~,
\label{T_I}
\eeqn
on the state-operator $\hat f$. If we write
${\hat f}(\vec{x})$ i.e.  ${\hat f}$ is an ordered ``function'' of $x_1$
and $x_2$ the constraint~\rref{T_I} implies
\beqn
{\hat  f}(\vec{x}+2\pi \vec{N}_i) ={\hat f}(\vec{x})
\exp\left(- i\zeta_i+ \frac{i m}{n}(1+\frac{\kappa}{\theta})\vec{N}_i\times
  \vec{x}\right)~,
\label{q-trans}
\eeqn
These relations are exactly the defining relations of quantum
bundles as discussed
in~\cite{Ho:1998hq}-\cite{Brace:1999xz}. Note that using the the
mapping~\rref{STAR} from commutative functions to operators, any vector
bundle with transition functions as in~\rref{c-trans} can be mapped into a
quantum bundle whose sections satisfy~\rref{q-trans}.

\section{Landau Levels}
\label{LLL}

We conclude by determining the structure of the energy spectrum of the
particle. We can immediately see that the eigenvalues of the energy
are independent of both the noncommutativity parameter and the torus
periods. The Hamiltonian~\rref{Ham} has a harmonic oscillator structure.
Defining the ladder operators $a,a^\dagger$
\[
a=(P_1 + i \,{\rm sgn} ({\cal B}) P_2)/\sqrt{2|{\cal B}|}
\]
the Hamiltonian becomes
\[
H = |{\cal B}| (a^\dagger a +\frac{1}{2})~.
\]
Therefore, its energy levels are of the form $E_n = |{\cal B}| (n+\frac{1}{2})$
which are the usual Landau level eigenvalues. (By the standard argument,
there cannot be any levels in between these values, since they would
violate unitarity.) It remains to determine the degeneracy of these
levels. The set of all degenerate states at a given energy level can be
obtained by acting on any
representative state for each level with the set of all physical
operators commuting with the Hamiltonian. The operators ${\tilde V}_i$
defined as
\[
\left\{~
\begin{array}{ccc}
{\tilde V}_1&=& \exp\left(- \frac{i}{B} p_1 \right) U_2~, \\
{\tilde V}_2&=& \exp\left(- \frac{i}{B} p_2 \right) U_1^{-1}~,
\end{array}\right.
\]
commute with $P_i$ and they are the minimal complete set of such operators
(it can be seen that the only operator commuting with $H$ but not the
individual $P_i$ is $H$ itself.)
They are the same as the operators $W_i$ defined in section 3, but with
the minimal choice of exponents $m_1 = m_2 =1$ since we are not concerned with
their commutation properties with the $U_i$. They can also be thought of as the
operators $V_i$ on the plane, generated by the magnetic translations,
but raised
to a fractional power $1/{\tilde B}$ in order to make the physical coordinate
operators $U_i = e^{i x_i}$ appear. They satisfy the commutation relations
\[
{\tilde V}_1 {\tilde V}_2 = e^{i\omega} {\tilde V}_2 {\tilde V}_1~,
 \label{VV}
\]
where the phase $\omega$ is
\[
\omega = \frac{1-\theta {B}}{B} = \frac{1}{\tilde B} = 2\pi \frac{n}{m}~.
\]
This is a `clock and shift' algebra whose irreducible representations are
$|m|$-dimensional (since $m$ and $n$ are relatively prime). In fact,
each degenerate energy multiplet forms one such irreducible
representation. Otherwise, there should exist some operator commuting
with $H$ and mixing the different irreducible components, and thus not
belonging to the set generated by the ${\tilde V}_i$, which is not the case
since ${\tilde V}_1 , {\tilde V}_2$
generate all the commutants of $H$. Therefore, we conclude that the
degeneracy of each Landau level is $|m|$.

The result above can be understood in terms of the density of states
in each Landau level in the planar case. Using the density of
states~\cite{Nair:2000ii}
\[
\rho = \frac{1}{2\pi} \left| \frac{\cal B}{1-\Theta \cal B} \right| ~,
\]
we obtain the total number of states
\[
\rho A = 2\pi |\tilde{B}|= |m/n|~.
\]
Consistency of quantization on the torus requires that there be an integer
number of states per total torus area. We saw that the Hilbert space of the
problem for $A {\cal B} /2\pi = m/n$ corresponds to quantizing on a larger
torus of area $n A$. On that torus the number of states per Landau level
is $\rho nA = |m|$.

The above results hold for $\Theta {\cal B}=\theta B \neq 1$. For the
critical case
${\cal B} = \Theta^{-1}$, ${\tilde B} = \infty$, the above argument gives an
infinite degeneracy of states per Landau level. In fact, at the
critical value of ${\cal B}$, the representation of the physical
observables reduces  into the sum of an infinite number of irreducible
components of the reduced algebra~\rref{Crit}. The operators ${\tilde U}_i$
are superselected and there is nothing that could induce transitions
between states $|\vec{\phi}\rangle$ with different $\vec \phi$. Each
irreducible component is labeled by the eigenvalues of ${\tilde U}_i$
and has a unique state per Landau level.

Finally,  the algebra~\rref{VV} has only infinite-dimensional 
representations when $A {\cal B} / 2\pi$ is irrational, and the degeneracy
of each Landau level becomes infinite. The Hilbert space of the
noncommutative torus in this case is the same as the one of the
noncommutative plane, since there is no finite multiple of the torus
containing an integer number of states per Landau level.

\section*{Acknowledgments}
We would like to thank K.~Bering and V.P.~Nair for helpful discussions.
This work was supported in part by the~ U.S.~ Department~ of~ Energy~
under ~Contract Number DE-FG02-91ER40651-TASK B.

\end{document}